\documentclass[twocolumn,assymb,amsmath]{revtex4}
\usepackage{mathrsfs}
\usepackage{txfonts}
\usepackage{graphicx}
\usepackage{amssymb}
\usepackage{amsmath}
\usepackage{wasysym}
\newcommand{\Tr}{\text{Tr}}

\begin{document}

\title{Dynamical localization in non-Hermitian quasi-crystals}
\author{C. M. Dai$^{1}$\footnote{daicm@zstu.edu.cn}, Yunbo Zhang$^{1}$\footnote{ybzhang@zstu.edu.cn}, and Xuexi Yi$^{2}$\footnote{yixx@nenu.edu.cn}}

\affiliation{$^{1}$Key Laboratory of Optical Field Manipulation of Zhejiang Province and Physics Department of Zhejiang Sci-Tech University, Hangzhou 310018, China\\
$^{2}$Center for Quantum Sciences, Northeast Normal University, Changchun 130024, China}

\date{\today}

\begin{abstract}
We study the localization transition in periodically driven one-dimensional non-Hermitian lattices where the piece-wise two-step drive is constituted by uniform coherent tunneling and incommensurate onsite gain and loss. We find that the system can be in localized, delocalized, or mixed-phase depending on the driving frequency and the phase shift of complex potential. Two critical driving frequencies of the system are identified, the first one corresponds to the largest phase shift of the complex potential so that the quasi-energy spectrum is still real and all the states are extended, the second one corresponds to the disappear of full real spectrum, and very weak complex potential leads to the emergence of localized states when the driving frequency is lower than this critical frequency. In the high frequency limit, we find the critical phase shift that separates the two regions with respectively real and complex spectrum tends to a constant value that can be captured by an effective non-Hermitian Hamiltonian.
\end{abstract}

\maketitle

\section{Introduction}
Periodically driven systems can host various exotic phases that have been extensively studied for many years, such as the engineering of nontrivial topological band structure \cite{jotzu2014,aidelsburger2015}, discrete time crystal \cite{else2016,yao2017}, driving induced effective density-dependent tunneling and associate entropic self-localization \cite{mamaev2019}. Recently, the influence of external field to the behavior of disordered systems attract lots of attention \cite{morales2014,bairey2017,bordia2017,xu2020,decker2020}. It is shown that the periodic driving can modify the Anderson localization \cite{anderson1958} or the many-body localization \cite{nandkishore2015} in a nontrivial manner. For example, the conversion of localized states to extended states by resonantly coupling \cite{morales2014}, driving induced many-body localization \cite{bairey2017}, and the control of topological properties of a many-body localized system by periodic driving \cite{decker2020}. Meanwhile, the physics of non-Hermitian disorder system is another hot research field \cite{hatano1996,efetov1997,hamazaki2019,longhi2019,longhi2019b,jiang2019,chen2019,kawabata2020,liu2020}, where the non-Hermiticity is typically achieved by introducing asymmetric tunneling or gain and loss that originates from the exchange of particles or energy with an environment. Different from their Hermitian counterpart, the localization transition of non-Hermitian systems can happen even in one dimension \cite{hatano1996}, and the localization length of non-Hermitian system can be anisotropic, even unidirectional delocalization \cite{kawabata2020}. Besides the system with random disorder, the localization transition is also found in non-Hermitian quasi-crystals \cite{longhi2019}. For non-Hermitian quasiperiodic lattices with exponentially short-range hopping, there exists energy-dependent mobility edges which separate the localized and extend states and are associated with the real part of eigenenergies \cite{liu2020}. It is also demonstrated that the localization transition point for general non-Hermitian quasicrystals with a complex phase factor can be determined by the Lyapunov exponents of its dual Hermitian model \cite{liu2021}.

However, little is known how the periodic driving changes the localization properties of non-Hermitian systems and the related real-complex spectrum transition \cite{hamazaki2019}. Here, we introduce a periodic driven non-Hermitian Aubry-Andr\'{e}-Harper (AAH) system and the piece-wise two-step driving constituted by uniform coherent tunneling and incommensurate onsite amplification and dissipation processes. We study the localization feature of this model when the driving frequency and strength changes, and identify the appearance of multiple phases for a broad range of driving parameters.

The paper is organized as follows. In Sec. \ref{model} we introduce the model. In Sec. \ref{fla} we present the basics of Floquet systems and calculate the effective Hamiltonian of our model. In Sec. \ref{qesl} we investigate the quasi-energy spectrum and localization transition of the driven system. In Sec. \ref{dyn} we study the Loschmidt echo dynamics for system in different phases. Finally, we conclude in Sec.\ref{con}.

\section{Model}\label{model}
As a starting point, we consider a model of non-Hermitian quasi-crystal in the presence of periodic drive, described by
\begin{equation}
H(t)=\left\{
\begin{aligned}
H_{A}\quad& mT\leq t<(m+1/2)T \\
H_{B}\quad& (m+1/2)T\leq t<(m+1)T,
\end{aligned}
\right.
\end{equation}
where $T$ is the period of drive, and $m$ is non-negative integer. In the following, we denote the frequency of drive by $\omega$, i.e. $\omega=2\pi/T$. The two parts $H_A$ and $H_B$ of this piece-wise drive are defined by
\begin{equation}
\begin{aligned}
H_{A}=& J \sum_n (|n\rangle \langle n+1|+|n+1\rangle \langle n|),\\
H_{B}=&  \sum_n V_n |n\rangle \langle n|,\\
\end{aligned}
\end{equation}
where $J$ is the tunneling amplitude, and the on-site potential takes form $V_n=V\cos(2\pi \alpha n+\phi)$ with strength $V$, incommensurate ratio $\alpha$ and phase shift $\phi$. The on-site potential $V_n$ is incommensurate with the lattice spacing when $\alpha$ takes a irrational number, and this is the case we will study in this work. The phase shift $\phi=\theta+i h$ is a complex number such that the on-site potential also becomes complex, except that for $h=0$ the Hamiltonian $H_B$ is Hermitian. We set incommensurate ratio as the inverse golden ratio $\alpha=(\sqrt{5}-1)/2$ in the following study.

\section{Floquet analysis}\label{fla}
One useful tool to analyze the properties of periodically driven systems is Floquet theorem \cite{goldman2014,bukov2015}. According to this theorem, the evolution operator of a periodically driven system, generated by time-periodic Hamiltonian $H(t+T)=H(t)$, can be written as (we set $\hbar=1$)
\begin{equation}
U(t)=P(t)\exp(-iH_F t),
\end{equation}
where the operator $P(t)$ is time-periodic $P(t+T)=P(t)$ with $P(0)=1$ being the identity, and the time-independent effective Hamiltonian $H_F$ is defined by the evolution operator over one driving period,
\begin{equation}\label{deff}
\exp(-iH_F T)\equiv\mathcal{T} \exp[-i\int_0^T d\tau H(\tau)].
\end{equation} 
The right side of this equation is just the formal solution of $U(T)$, and $\mathcal{T}$ represents the time-ordering operator. The coarse grained evolution of system over many periods $mT$ can be given by the effective Hamiltonian $H_F$ or $U(T)$, i.e.
\begin{equation}
U(mT)=U^m(T)=\exp(-im H_F T).
\end{equation}
The eigenvectors $|\lambda \rangle$ of $U(T)$ are called Floquet modes, where we denote the corresponding eigenvalues by $\mu_\lambda$ with label $\lambda$. If $H(t)$ is Hermitian at time $t$, the evolution operator $U(T)$ is a unitary operator, all the eigenvalues $\mu_\lambda$ of $U(T)$ lie on the unit circle, and the effective Hamiltonian $H_F$ can also be chosen to be Hermitian. When we consider non-Hermitian Hamiltonian $H(t)$, the norm of the eigenvalues of $U(T)$ can be smaller or larger than unity, and the corresponding quasi-energy $\varepsilon_\lambda=i\ln(\mu_\lambda)/T$ may have a nonzero imaginary part. In this case, we denote the real and imaginary parts of $\varepsilon_\lambda$ by $\varepsilon_\lambda^{R}$ and $\varepsilon_\lambda^{I}$, respectively, i.e. $\varepsilon_\lambda = \varepsilon_\lambda^R+i \varepsilon_\lambda^I$. It may be useful to note that $\det U(T)=\exp[-i\int_{0}^{T} d\tau \Tr H(\tau)]$ \cite{blanes2009}, and in our case with the so called periodic boundary condition \cite{longhi2019}, we have $\det U(T) = 1 =\prod_\lambda \mu_\lambda$. In the following we are interested in the bulk properties of $H(t)$, and always assume periodic boundary condition in the calculations, the equation $1 =\prod_\lambda \mu_\lambda$ implies that either all the quasi-energies are real or quasi-energies with positive and negative imaginary parts coexist.

When the frequency of drive is high, one typically can calculate the effective Hamiltonian $H_F$ approximately by the Magnus expansion \cite{blanes2009,bukov2015} or Baker-Campbell-Hausdorff (BCH) formula \cite{blanes2009} for the piece-wise drive that we consider in this work. BCH formula gives a series expansion of $Z$ in the following equation
\begin{equation}
\exp(Z)=\exp(X)\exp(Y),
\end{equation}
i.e. $Z$ can be written as a series
\begin{equation}\label{bch}
Z=\sum_{n=1}^\infty z_n(X,Y),
\end{equation}
where $z_n(X,Y)$ represents homogeneous Lie polynomial in $X$ and $Y$ of grade $n$ \cite{blanes2009}. The first two terms in the summation of Eq.(\ref{bch}) have relatively simple forms
\begin{equation}\label{fwe}
\begin{aligned}
z_1(X,Y)&=X+Y,\\
z_2(X,Y)&=\frac{1}{2}[X,Y].
\end{aligned}
\end{equation}
With the definition Eq.(\ref{deff}) of the effective Hamiltonian $H_F$ and the explicit expression Eq.(\ref{fwe}) of the first two order expansions, the approximate effective Hamiltonian $H_F$ up to the second order can be written as
\begin{equation}
H_F\approx H_{F1}+H_{F2},
\end{equation}
with
\begin{equation}
\begin{aligned}
H_{F1}&=\frac{1}{2}(H_A+H_B),\\
H_{F2}&=\frac{i\pi J}{4\omega}\sum_n (V_{n+1}-V_n)(|n\rangle \langle n+1|-|n+1\rangle \langle n|),
\end{aligned}
\end{equation}
Here $H_{F1}$ is just the non-Hermitian extension of the AAH model that has been studied recently \cite{longhi2019,longhi2019b,liu2020}. The non-Hermitian AAH model undergoes a phase transition at $h_c=\ln(2J/V)$, i.e. the energy spectrum changes from entirely real to complex when the imaginary part of the phase shift is greater than the critical value $h_c$. Accompanied by this phase transition, all the eigenstates of $H_{F1}$ become localized for $h>h_c$. The second order correction $H_{F2}$ is proportional to the inverse of driving frequency $1/\omega$ that represents the leading order correction to the effective Hamiltonian at finite driving frequency, and the effect of $H_{F2}$ will be discussed in detail in the following section.

\section{Quasi-energy spectrum and localization transition}\label{qesl}
\begin{figure}
	\includegraphics[scale=0.88]{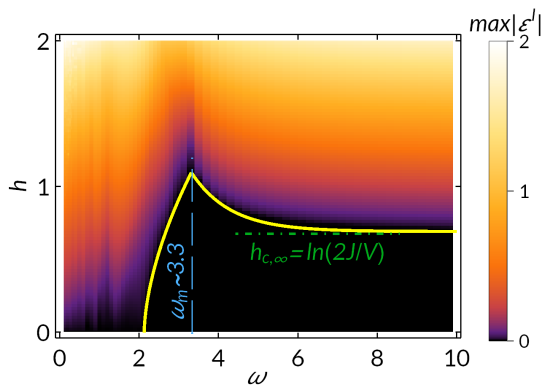}
	\caption{The largest value of $|\varepsilon^I|$ versus complex phase shift $h$ and driving frequency $\omega$ for $J=V=1$, $\alpha=(\sqrt{5}-1)/2$, $\theta=0$, and the length of lattice $L=610$. The yellow line separates real and complex regions of the quasi-energy $\varepsilon$. The dot dashed line represents the critical value of phase shift in the high frequency limit $h_{c,\infty}$ that separates two different phases. When the frequency $\omega$ decrease, the frequency dependent critical non-Hermitian phase $h_{c,\omega}$ gradually increases to its maximum with $\omega_m\approx3.3$ (blue dashed line) then quickly decreases to zero near $\omega_c\approx2.1$.}\label{eg}
\end{figure}

First, we study the appearance of a nonzero imaginary part of the quasi-energy for general driving parameters. We calculate the quasi-energy spectrum of $H_F$ by numerically diagonalizing the time evolution operator $U(T)$ under periodic boundary condition, and show the largest $|\varepsilon^I|$ as function of complex phase shift $h$ and driving frequency $\omega$ for $J=V=1$ in Fig.\ref{eg}. We find that as suggested by the approximate effective Hamiltonian $H_{F1}$, a nonzero imaginary part of the quasi-energy appears when $h>h_{c,\infty}=\ln(2J/V)$ for sufficiently high driving frequency $\omega\sim10$, here and following we denote the frequency dependent critical complex phase shift by $h_{c,\omega}$. 

\begin{figure}
	\includegraphics[scale=0.63]{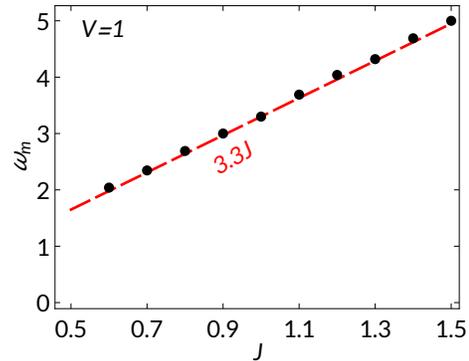}
	\caption{The driving frequency $\omega_m$ that critical complex phase shift $h_{c,\omega}$ reaches its maximum value versus different tunneling strength $J$. The red dashed line is the linear fitting $\omega_m \sim 3.3J$. The other parameters used in the calculation $V=1$, $\alpha=(\sqrt{5}-1)/2$, $\theta=0$, and the length of lattice $L=610$.}\label{wm}
\end{figure}

If we gradually decrease the driving frequency $\omega$, the complex phase shift $h_{c,\omega}$ that governs the appearance of complex quasi-energy gradually increases to its maximum value when $\omega$ approaches $\omega_m\approx3.3$. We find that the driving frequency $\omega_m$ corresponding to the largest $h_{c,m}$ has a quiet simple dependence on the tunneling strength $J$, i.e. $\omega_m = \kappa J$ where the scaling factor $\kappa \approx 3.3$ are obtained by fitting the numerical solution of $\omega_m$ with different $J$. We show this in Fig.\ref{wm}, where the black dots are the results of $\omega_m$ for different tunneling strength $J$ by direct diagonalizing Floquet propagator $U(T)$, and the red dashed line is the linear fitting $\omega_m = \kappa J$ that agrees well with the numerical exact results. For sufficiently slow drive $\omega<\omega_c\approx2.1$ shown in the left side of Fig.\ref{eg}, the critical complex phase shift $h_{c,\omega}$ is close to zero that means very weak non-Hermitian perturbation characterized by complex phase shift $h$ can introduce a non-vanishing imaginary part to the quasi-energy. Note that the similar results can be found for other irrational incommensurate ratio $\alpha$, though the scaling factor $\kappa$ and the value of $\omega_c$ may be slightly different, for example, when $\alpha=\sqrt{3}/3$, we find $\kappa\approx3.6$ and $\omega_c\approx2.4$, respectively, and the results are insensitive to specific choice of the real phase $\theta$.

The largest value of $|\varepsilon^I|$ shows rich behavior along with the increasing of driving frequency $\omega$, and one may expect the effective Hamiltonian $H_F\approx H_{F1}+H_{F2}$ with finite frequency correction $H_{F2}\propto 1/\omega$ would give us similar $max|\varepsilon^I|$ at a relative large but finite driving frequency comparing with the exact results shown in Fig.\ref{eg}, as what happened in Hermitian systems \cite{kitagawa2011,dai2018}. Afterall $h_{c,\omega}$ tends to a high frequency limit value for large $\omega$. However we find that the finite frequency correction $H_{F2}$ is not good enough to characterize the change from entirely real to complex quasi-energy spectrum in the non-Hermitian Floquet system studied here.

\begin{figure}
	\includegraphics[scale=0.65]{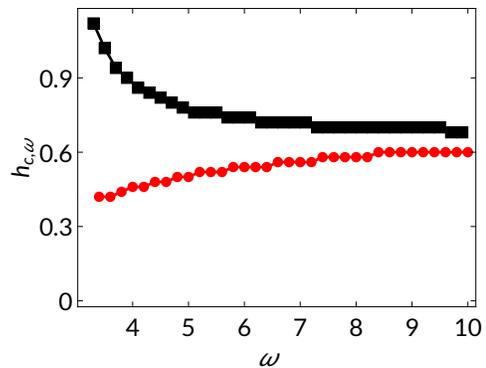}
	\caption{Comparison between the results of frequency dependent critical complex phase shift $h_{c,\omega}$ calculated by the exact Floquet evolution operator $U(T)$ and the approximated effective Hamiltonian $H_F=H_{F1}+H_{F2}$, square and dot symbols for $U(T)$ and $H_F$, respectively. The parameters used in the calculation $J=V=1$, $\alpha=(\sqrt{5}-1)/2$, $\theta=0$, and the length of lattice $L=610$.}\label{cp}
\end{figure}

To clarify this, we plot the frequency dependent critical complex phase shift $h_{c,\omega}$ calculated by diagonalizing both $H_F\approx H_{F1}+H_{F2}$ and $U(T)$ in Fig.\ref{cp}. When the driving frequency $\omega\sim 10$ is one order of magnitude larger than $J$ and $V$, the difference between the two results is still relatively large, and the approximate effective Hamiltonian $H_F\approx H_{F1}+H_{F2}$ predicts a decreasing trend of $h_{c,\omega}$ along with the decreasing of driving frequency $\omega$ that is in contrast with the result given by $U(T)$. Note that the deviation of the results calculated by the approximate $H_F$ and $U(T)$ is smaller for high frequency drive when the third order correction $H_{F3}\propto1/\omega^2$ is included, which, however still can not give the correct variation tendency of $h_{c,\omega}$ with the decreasing of driving frequency $\omega$.

\begin{figure}
	\includegraphics[scale=0.57]{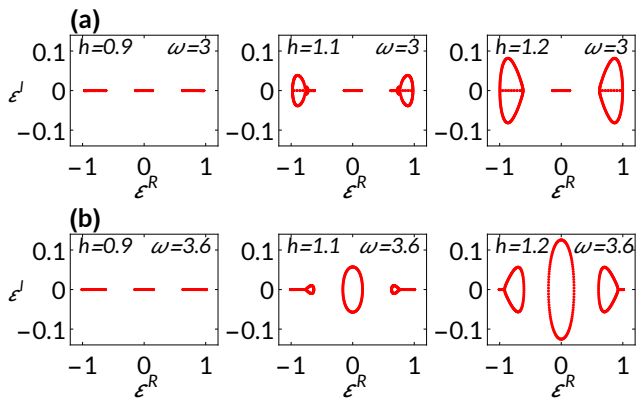}
	\caption{Quasi-energy spectrum for a few increasing complex phase shift $h$. In (a) and (b) we choose two different driving frequencies $\omega=3$ and $3.6$ that are on the left and right sides of the critical frequency $\omega_m\approx3.3$, respectively. The other parameters $J=V=1$, $\alpha=(\sqrt{5}-1)/2$, $\theta=0$, and the length of lattice $L=610$.}\label{tt}
\end{figure}

Different from the original non-Hermitian AAH model that almost all (with a few exceptions) eigenenergies become complex when $h>h_c$ \cite{longhi2019}, for the periodic driven case, there can be only a part of eigenstates acquire complex (quasi)energies when $h>h_{c,\omega}$. This can be seen clearly in Fig.\ref{tt} (a) and (b) that show the quasi-energy spectrum for two different driving frequency. The driving frequency $\omega$ used in Fig.\ref{tt} (a) is smaller than $\omega_m$, and $\omega$ is larger than $\omega_m$ in Fig.\ref{tt} (b). Fig.\ref{tt} (b) shows that for driving frequency $\omega=3.6>\omega_m$, the states in the central band acquires complex energy prior to the other bands with the increasing of the strength of complex phase shift $h$, and the situation is just the opposite when $\omega=3.0<\omega_m$ shown in Fig.\ref{tt} (a). In the latter case as shown in the right of Fig.\ref{tt} (a), though most of the quasi-energies in the left and right bands are complex, a fraction of quasi-energies can still be real, and the complex quasi-energies form a closed curve that encircles these real quasi-energies.

With the quasi-energy spectrum of the periodically driven system in hand, next we study the localization transition of the driven system and the relevance between the localization transition and the nature of the quasi-energy spectrum. The quantity used here to quantify the localization degree of a state $|\psi\rangle=\sum_i \psi_i |i\rangle$ is the inverse participation ratio (IPR)
\begin{equation}
IPR=\sum_i |\psi_i|^4,
\end{equation}
where $\psi_i$ is the normalized complex amplitude at site $i$, to be specific $\sum_i |\psi_i|^2=1$. For a state fully localized at some sites, $IPR=1$, and for a state distributes uniformly over the lattice $IPR=1/L$ where $L$ is the length of lattice. Generally, for localized states the inverse participation ratio $IPR\sim1$ is finite and for extended states $IPR\sim1/L$ vanishes when the length $L$ of lattice tends to infinity.

\begin{figure}
	\includegraphics[scale=0.63]{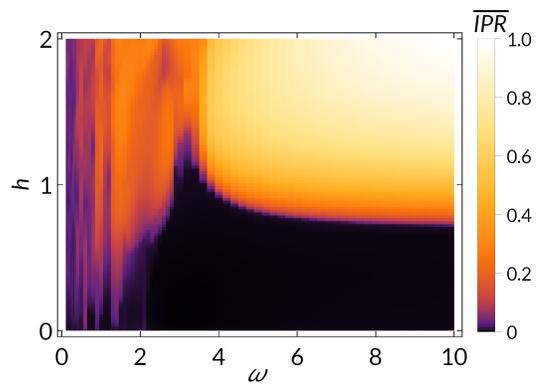}
	\caption{The average inverse participation ratio over all eigenstates of $U(T)$ versus complex phase shift $h$ and driving frequency $\omega$ for $J=V=1$, $\alpha=(\sqrt{5}-1)/2$, $\theta=0$, and the length of lattice $L=610$.}\label{ave}
\end{figure}

To gain some insight of the localization properties of the driven system, we calculate the average IPR over all the eigenstates of system for various driving frequency $\omega$ and complex phase shift $h$ and the results are shown in Fig.\ref{ave}. The dark region in Fig.\ref{ave} that suggests the delocalization of states is in agreement with the region of entirely real spectrum shown in Fig.\ref{eg}. The behavior of average IPR along with the changing of $h$ is similar with the original non-Hermitian AAH model when the driving frequency $\omega$ is larger and not too close to $\omega_m$. As for the low frequency driving $\omega<\omega_m$, the behavior of average IPR is not so regular as the high frequency one. 

\begin{figure}
	\includegraphics[scale=0.6]{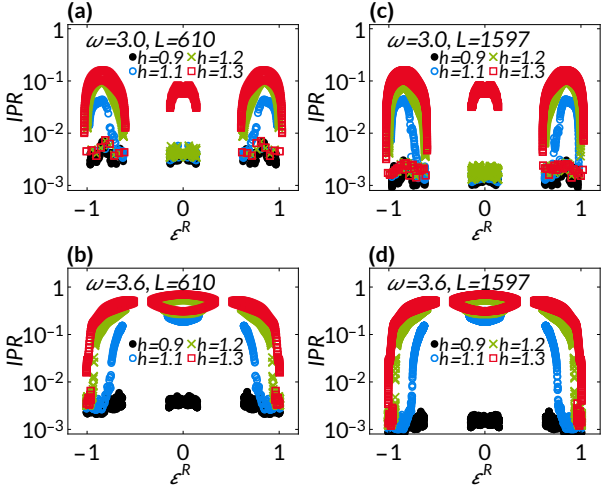}
	\caption{(a) and (b) The inverse participation ratio versus the real part of quasi-energy for a few different complex phase shift $h$ and driving frequency $\omega$. The other parameters used in the calculations $J=V=1$, $\alpha=(\sqrt{5}-1)/2$, $\theta=0$, and the length of lattice $L=610$. (c) and (d) are the same as (a) and (b), except the length of lattice $L=1597$.}\label{iprt}
\end{figure}

\begin{figure}
	\includegraphics[scale=0.73]{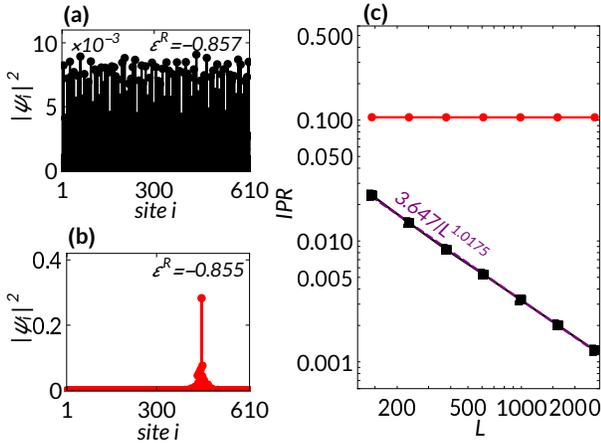}
	\caption{(a) and (b) The different density distributions of the eigenstates of $U(T)$ with very close $\varepsilon^R$, the other parameters  $J=V=1$, $\alpha=(\sqrt{5}-1)/2$, $\theta=0$, $\omega=3.0$, $h=1.3$, and the length of lattice $L=610$. (c) The scaling of $IPR$ as functions of lattice length $L$ for the first two states with $\varepsilon^R$ close to $-0.856$. The dashed line indicates the power law fitting.}\label{wav}
\end{figure}

As suggested by the coincidence of non-Hermitian localization transition with the real-complex transition of spectrum \cite{longhi2019,liu2020} and the partial real quasi-energy spectrum of our system for relative low frequency driving shown in Fig.\ref{tt}, one may expect the coexistence of localized and extended states in the driven system, and this is indeed the case. Fig.\ref{iprt} shows the IPR versus the real part of quasi-energy for two different driving frequency $\omega=3.0,3.6$ and a few increasing complex phase shift $h$. Comparing with the case of relative short
lattice length $L=610$ shown in Fig.\ref{iprt} (a) and (b), the IPR versus $\varepsilon_R$ for $L=1597$ plotted in Fig.\ref{iprt} (c) and (d) shows  similar structure where the data with larger IPR is almost the same, except the points are more dense, and those near zero becomes closer to zero. Except for the coexistence of localized and extended states, the results in Fig.\ref{iprt} also indicate that the states with similar $\varepsilon^R$ can display very different localization properties. It is easier to see this by taking two eigenstates as examples in the left side of Fig.\ref{iprt} (a) with $-1<\varepsilon^R<-0.5$. We plot the spatial distributions of two eigenstates with very close $\varepsilon^R=-0.855$ and $-0.857$ but drastically different localization properties for the case $h=1.3$ shown in Fig.\ref{iprt} (a) in Fig.\ref{wav} (a) and (b) as an illustration. Fig.\ref{wav} (c) shows how the $IPR$ of such states in Fig.\ref{wav} (a) and (b) changes as functions of lattice length $L$, we can see that $IPR$ is either close to a constant or inversely proportional to the lattice length that indicates one kind of state is localized (dot line) and another (square line) is extended, respectively. In Hermitian quasiperiodic lattices, localized and extended states can also coexist, the values of energy that separate the states with different localization properties are called mobility edges \cite{sarma1990,biddle2010,xu2019,ganeshan2015,wang2020}. There are also such mixed-phase in non-Hermitian systems \cite{liu2020}, but it is different from the previous work that considers static non-Hermitian system, here we find that the mixed-phase can emerge in periodically driven non-Hermitian system when the driving is relatively slow.

\section{Dynamics}\label{dyn}
Besides the properties of the Floquet modes, the localization property can also be probed by the dynamics of system. Here, we are interested in the Loschmidt echo dynamics of the periodically driven non-Hermitian system, recent works show that the Loschmidt echo dynamics can characterize the localization transition in both Hermitian and non-Hermitian generalization of AAH models \cite{yang2017,liu2020}. The Loschmidt echo is defined by the overlap of an initial state $|\psi(0)\rangle$ with its post-quench state $|\psi(t)\rangle=U(t)|\psi(0)\rangle$, i.e.
\begin{eqnarray}
L(t)=|\langle \psi(0)|\psi(t)\rangle|^2 / (\langle \psi(t)|\psi(t)\rangle \langle \psi(0)|\psi(0)\rangle).
\end{eqnarray}

\begin{figure}
	\includegraphics[scale=0.55]{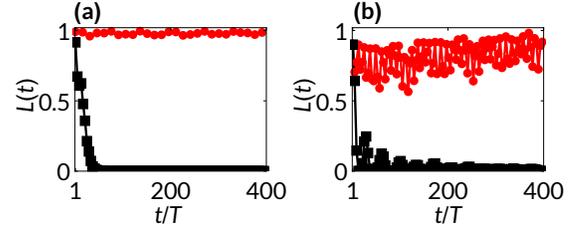}
	\caption{Loschmidt echo as functions of evolution time. (a) The initial states are the eigenstates of the the system in the extend regime with $h=0.8$ and minimum eigenvalue, final system in the extended and mixed regimes with $h=0.9$ and $h=1.1$ are denoted by dots and squares, respectively. (b) The initial states are the eigenstates of the the system in the mixed regime with $h=1.1$, dot line for initial state with minimum $\varepsilon^R$ that is extended, square line for localized initial state with $\varepsilon^R=0$, and the finial system is in the extended regime with $h=0.8$.}\label{l}
\end{figure}

We plot the time evolution of Loschmidt echo of our system in Fig.\ref{l}. Fig.\ref{l} (a) shows the quench dynamics for system initially prepared in the eigenstate of system in the extended regime and the final systems in the extend and mixed regimes, respectively. When the final system in the extended regime, the Loschmidt echo oscillates around a value near unity, but for final system in the mixed regime, the Loschmidt echo decreases to zero after a short transition time. Fig.\ref{l} (b) shows the quench dynamics from the mixed regimes to extended regime, with two different choices of initial states. The dot line with extended initial state shows that the evolution of Loschmidt echo is similar with the quench dynamics from extended to extended regime shown in Fig.\ref{l} (a), while the square line represents quench dynamics of initially localized states where the Loschmidt echo approaches zero at long time though oscillates at short time. The above observations agree with the time evolution of Loschmidt echo for systems with initial states in different regimes studied in Ref.\cite{liu2020}.

\section{Conclusion}\label{con}
In this work we have studied the localization transition of a non-Hermitian system with piece-wise periodic drive composed by uniform tunneling and quasi-periodic complex onsite potential. We show that for high frequency drive, the localization properties of our system is similar with the non-Hermitian extension of AAH model studied recently, i.e. all the eigenstates change from being extended to localized when the strength of the non-Hermitian phase shift is larger than a specific value. If we slightly lower the driving frequency, the critical non-Hermitian phase shift needed for the appearance of localized states increases, until the driving frequency reaches a bound that is proportional to the tunneling strength. Near this bound, the critical non-Hermitian phase shift takes its maximum value, and we find the coexistence of localized and extended states in the system. Further lower the driving frequency, we find very weak non-Hermitian phase shift can localize the states of system, which means that the extended phase is more fragile with slow driving in the periodically driven  non-Hermitian quasi-periodic system. We also study the Loschmidt echo dynamics of our system that agrees with the recent work \cite{liu2020} for effective Floquet system in the different regimes.

\section*{ACKNOWLEDGMENTS}
This work is supported by National Natural Science Foundation of China (NSFC) under Grants No. 12105245, No. 12074340, No. 11775048.

\end{document}